\documentstyle [fleqn,12pt]{article}
\begin{document}
\baselineskip .75cm 
\begin{titlepage}
\title{ \bf Single parameter quasi-particle model for QGP}      
\author{Vishnu M. Bannur \\
{\it Department of Physics}, \\  
{\it University of Calicut, Kerala-673 635, India.} }
%\date{}
\maketitle
\begin{abstract} 

We discuss a new single parameter quasi-particle model and study the 
thermodynamics of (2+1)-flavor quark gluon plasma (QGP). Our model 
with a single parameter explains remarkably well the lattice simulation 
results of Fodor {\it et. al.} \cite{fk.1}.  
\end{abstract}
\vspace{1cm}

\noindent
{\bf PACS Nos :} 12.38.Mh, 12.38.Gc, 05.70.Ce, 52.25.Kn \\
{\bf Keywords :} Equation of state, quark gluon plasma, 
quasi-particle plasma  
\end{titlepage}
%%%%%%%%%%%%%
\section{Introduction :}

Quasi-particle model (qQGP) of quark gluon plasma (QGP) was first proposed by 
Peshier {\it et. al.} \cite{pe.1} to explain the non-ideal equation of state 
(EoS), observed in lattice gauge theory simulations (LGT) \cite{rh.1}. 
At finite temperature, instead of 
real quarks and gluons {\it with} QCD (quantumchromodynamics) 
{\it interactions} we may as well consider the system to be made up of 
{\it non-interacting} quasi-particles with thermal masses, quasi-quarks 
and quasi-gluons, and study the thermodynamics. Quasi-particles are quanta 
of plasma collective modes excited by quarks and gluons through QCD 
interactions. Initial quasi-particle model was found to be 
thermodynamically inconsistent \cite{go.1} and also not able to fit 
the more recent LGT results \cite{ba.1}. Gorenstein and Yang 
reformulated the statistical mechanics (SM) to solve the inconsistency, 
but end up with an extra undetermined, temperature dependent terms in 
the expressions for pressure and energy density, which need to be 
phenomenologically chosen. However, as we discuss here \cite{ba.2}, 
above reformulation of SM is not needed and standard SM may be applied 
to qQGP without any extra phenomenological terms. There is no 
TD inconsistency in our new qQGP model. Peshier's model with reformulated 
SM by Gorenstein and Yang  has been studied by various groups 
\cite{pe.2,lh.1,s.1,st.1,is.1} with different expressions  
for thermal masses, effective degrees of freedom, so on. Note that 
all of above works are based on the reformulation of SM of Gorenstein and Yang, 
which in fact based on mathematical identities involving derivatives with 
respect to temperature and chemical potentials, used to redefine average 
energy density ($\varepsilon$) and number density ($n$) respectively. 
As we have shown recently in Ref. \cite{ba.2}, we may skip this TD 
inconsistency problem and instead  use the original definition 
of $\varepsilon$ and $n$, and making use of TD relations we may get all 
TD quantities. As a specific example, here we discuss (2+1)-flavor QGP, 
studied by Fodor {\it et. al.} \cite{fk.1} using LGT. 

Of course, there are other models like HTL (hard thermal loop) \cite{bi.1},  
recent FMR (fundamental modular region) gas \cite{d.1} etc. based on 
QCD perturbative and non-perturbative calculations, but fails to 
fit LGT results near to the transition temperature $T_c$. At the same 
time phenomenological models of QGP, based on plasma theory with QCD 
inputs like SCQGP (strongly coupled quark gluon plasma) \cite{ba.3}, 
our present qQGP model seems to fit remarkably well the LGT results with 
minimum number of parameters. All other qQGP models, field theoretical 
models \cite{rt.1}, Ploykov loop models \cite{rh.2} also fit LGT results, 
but by adjusting $3$ or more parameters. 
Of course, we know that near $T =  T_c$, region of phase transition
or cross over, it is a low energy phenomena and hence QCD can not be solved 
by analytical methods like perturbation theory because the coupling 
constant $\alpha_s$ is not small enough. Probably we need to formulate  
phenomenological models, just like in the case of hadron spectroscopy, to 
study TD of QGP near $T_C$. 

\section{Phenomenological Model with $\mu = 0$:}

QGP at thermodynamic equilibrium consists of interacting quarks and gluons 
which exhibits collective behaviour. Our basic assumption is that 
this system may be replaced by a system of non-interacting quasi-particles 
with quantum numbers of quarks and gluons. These quasi-particles have 
additional thermal masses which are equal to plasma frequencies. Here we 
differ from other qQGP models where, for example, the thermal mass was 
taken to be $\sqrt{3/2}$ times the plasma frequency. A general expression for 
thermal mass or polarization tensor is very complicated expressions 
which is a function of momentum and frequency. Only at high momentum 
limit it approaches a simpler form which on further approximations 
reduces to above form. In view of such a drastic approximation and since we  
use phenomenological model one may as well take $m_{th} \approx \omega_p$. 
In fact with this relation, we get better result than with  
$m_{th} \approx \sqrt{3/2} \,\omega_p$. Further important point is that 
the above dispersion relation is obtained using perturbation methods 
with temperature dependent density distribution function appropriate 
to ideal system. Then one formulates TD of a non-ideal system.  
In principle, this must be carried out in a self-consistent manner as 
discussed in Ref. \cite{ba.4}, where, for example, density expression 
is an integral equation since $\omega_p$ depends on density. So we need 
to solve an integral equation self-consistently to get the   
the density. Here we avoid all above complications and as a 
phenomenological input, we assume that $m_{th} = \omega_p$.  

Following the standard procedure of statistical mechanics \cite{pa.1}, 
the grand partition function is given by,  
\begin{equation}
Q_G = \sum_{s,r} e^{- \beta E_r  - \alpha N_s}
\,\,, \end{equation}
where the sum is over energy states $E_r$ and particle number states $N_s$. 
$\alpha$ and $\beta$ are defined as $\alpha \equiv -\mu /T$ and 
$\beta \equiv 1/T$.     
Next on further simplifications and taking thermodynamic limit, we get,  
\begin{equation} 
q \equiv \ln Q_G = \mp \sum_{k=0}^\infty 
\ln (1 \mp z\, e^{ - \beta \epsilon_k})  
\,\,, \label{eq:q} \end{equation} 
where $q$ is called q-potential and $\mp$ for bosons and fermions. 
$z \equiv e^{\mu /T} = e^{-\alpha}$ is called fugacity. $\epsilon_k$ 
is the single particle energy, given by, 
\[ \epsilon_k = \sqrt{k^2 + m^2 } \,\,, \]   
where $k$ is momentum and $m$ is the total mass which contains 
both the rest mass and thermal mass ($m_{th}$) of particles. 
$m_{th}$ may depend on temperature $T$ and chemical potential $\mu$ 
depending on QGP system.  
All other qQGP models starts from q-potential to get pressure with correction 
term to avoid TD inconsistency and proceed to evaluate other TD quantities. 
However, we avoid this trap of inconsistency problem by adopting a different 
procedure. We know that, for grand canonical ensemble, TD quantities like 
energy density and number density may be obtained by taking ensemble averages.  
That is, the average energy $U$ is given by, 
\begin{equation}
U \equiv <E_r> = \frac{\sum_{s,r}\,E_r \,e^{- \beta E_r - \alpha N_s}}{Q_G} = 
- \frac{\partial }{\partial \beta} \ln Q_G = 
\sum_k \frac{z \,\epsilon_k e^{- \beta \epsilon_k }}{1 \mp 
z\, e^{- \beta \epsilon_k} } 
\,\,.\end{equation} 
Note that the partial differentiation with respect to $\beta$ above is just 
a mathematical trick to express $U$ in terms of sum over single particle 
energy levels, $\epsilon_k$, making use of Eq. (\ref{eq:q}). While 
differentiating, indirect dependence of $\beta = 1/T$ in the fugacity, $z$, 
and thermal mass, $m_{th}(T,\mu)$, must be ignored by definition. 
Otherwise, we will not get back $<E_r>$.    
Now, on taking continum limit and after some algebra, we get, 
\begin{equation}
\varepsilon = \frac{g_f\, T^4}{2\, \pi ^2} \sum_{l=1}^\infty 
(\pm 1)^{l-1} z^l \frac{1}{l^4} \left[ (\frac{m\,l}{T})^3 K_1 (\frac{m\,l}{T}) 
+  3\, (\frac{m\,l}{T})^2 K_2 (\frac{m\,l}{T}) \right] 
\,\,, \label{eq:u} \end{equation}
where $g_f$ is the degenarcy and equal to $g_g \equiv 16$ for gluons and 
equal to $12\,n_f$ for quarks. $n_f$ is the number of flavors with same mass. 
$K_1$ and  $K_2$ are modified Bessel functions of order 1 and 2 respectively. 

Let us now consider our main topic, 
(2+1)-flavor system, studied by Fodor {\it et. al.} \cite{fk.1}, 
using our model. It is a QGP with two light ($u$) 
and one heavy ($s$) quarks along with gluons. 
Let us first consider the case with zero chemical potential and take $z=1$.
Hence we get the energy density, expressed in terms of
$e(T) \equiv \varepsilon / \varepsilon_s$,
for the quark gluon plasma of quasi-partons is
\[ e(T) = \frac{15}{\pi ^4} \frac{1}{(g_f + \frac{21}{2} n_f^{eff})}
\sum_{l=1}^\infty  \frac{1}{l^4}  \left( g_f \,
\left[ (\frac{m_g\,l}{T})^3 K_1 (\frac{m_g\,l}{T})
+  3\, (\frac{m_g\,l}{T})^2 K_2 (\frac{m_g\,l}{T}) \right] \right. \]
\[
 + 12\,n_u^{eff}\,
(-1)^{l-1}  \left. \left[ (\frac{m_u\,l}{T})^3 K_1 (\frac{m_u\,l}{T})
+  3\, (\frac{m_u\,l}{T})^2 K_2 (\frac{m_u\,l}{T}) \right] \right.  
\,\, \]
\begin{equation}
 + 12\,n_s^{eff}\,
(-1)^{l-1}  \left. \left[ (\frac{m_s\,l}{T})^3 K_1 (\frac{m_s\,l}{T})
+  3\, (\frac{m_s\,l}{T})^2 K_2 (\frac{m_s\,l}{T}) \right] \right)
\,\,, \end{equation}
where $\varepsilon_s$ is the Stefan-Boltzman gas limit of QGP, which
may be obtained by taking high temperature limits of Eq. (\ref{eq:u})
for gluons and quarks separately and adding them.
$m_g$ is the temperature dependent gluon mass ($m_{th}$), which is equal to the
plasma frequency, i.e,
$m_g^2 = \omega_p^2 = \frac{g^2 T^2}{18} (2 N_c + n_f)$. 
All quarks have both the thermal mass as well as the rest mass and hence, 
the total mass may be written as    
\begin{equation}
m_q^2 = m_{q0}^2 + \sqrt{2} \,m_{q0}\,m_{th} + m_{th}^2\,\,,  
\end{equation}
following the idea used in other qQGP models for the system with finite 
quark masses. Only the difference is that our $m_{th}$ is equal to the 
plasma frequency due quarks alone. That is, 
$m_{th}^2 = \omega_p^2 = \frac{g^2 T^2}{18} \,n_f$. 
$m_{q0}$ is the rest mass of up or strange quark. 
$g^2$ in thermal masses is related to the two-loop order running 
coupling constant, given by,
\begin{equation} \alpha_s (T) = \frac{6 \pi}{(33-2 n_f) \ln (T/\Lambda_T)}
\left( 1 - \frac{3 (153 - 19 n_f)}{(33 - 2 n_f)^2}
\frac{\ln (2 \ln (T/\Lambda_T))}{\ln (T/\Lambda_T)}
\right)  \label{eq:ls} \;, \end{equation}
where $\Lambda_T$ is a parameter related to QCD scale parameter. This
choice of $\alpha_s (T)$ is motivated from lattice simulations.
Using thermal masses with above $\alpha_s$, we
can evaluate the $e(T)$ from Eq. (\ref{eq:u}). Note that the only
temperature dependence in $e(T)$ comes from $\alpha_s(T)$, which
has the same form as that of lattice simulations \cite{ka.1} with
$\Lambda_T$ as a free parameter. $n_u^{eff}$ and $n_s^{eff}$ are 
effective number of quarks which differ from $2$ and $1$, respectively, 
because of finite masses of up and strange quarks. However, handling of 
finite rest masses of quarks are different in LGT studies of 
Fodor {\it et. al.} and Bielefeld group \cite{ka.1}. Bielefeld group carried 
out the simulation with the ratio $m_{q0}/T$ equal to constant and it is 
straight forward to calculate $n_q^{eff}$, where as 
Fodor {\it et. al.} used constant $m_{q0}$ and it is not clear how to 
calculate $n_q^{eff}$. Hence as done in Ref. \cite{ra.2} we take in our 
calculations $n_u^{eff} = 2$, $n_s^{eff} = 0.5$ and $n_f^{eff} = 2.5$. 
Similar normalization  
need to be made to fit the LGT results in other models also,  
either multiplying LGT data with a factor $1.1$ \cite{s.1} or model's data 
with $.9$ \cite{is.1}, or sometime using massive gluon \cite{ra.1} so on.  
From $\varepsilon$, we may obtain pressure $P$ by using a TD relation 
$\varepsilon =  T \frac{\partial P}{\partial T} - P $ for $\mu =0$ system 
and we get
\begin{equation} \frac{P}{T} = \frac{P_0}{T_0} + \int_{T_0}^{T} dT \,
\frac{\varepsilon (T)}{T^2} \,\, , \label{eq:p} \end{equation}
where $P_0$ and $T_0$ are pressure and temperature at
some reference points. Results are presented in Fig. 1 along with 
LGT results. Note that earlier 
this phenomenological new qQGP model with a single system dependent 
adjustable parameter explained very well the LGT results of 
Bielefeld group \cite{ka.1} on QGP system with massless quarks 
as discussed in Ref. \cite{ba.2}. 

Another TD quantity which may be obtained from ensemble averaging is 
the number density, given by,    
\begin{equation}
<N> = \frac{\sum_{s,r}\,N_s \,e^{- \beta E_r - \alpha N_s}}{Q_G} = 
- \frac{\partial }{\partial \alpha} \ln Q_G = 
 z \frac{\partial }{\partial z} \ln Q_G =  
\sum_k \frac{z \,e^{- \beta \epsilon_k }}{1 \mp 
z\, e^{- \beta \epsilon_k} } \,\,, \label{eq:n} 
\end{equation}
and using the definition $n \equiv <N>/V$, where $V$ is the volume. 
 
\section{Model with finite $\mu$ :} 

Let us next consider (2+1)-flavor system with finite $\mu$ and 
LGT results are available for quark density $n_q$ or 
baryon density ($n_B$) and difference in pressure 
from $\mu=0$ case ($\Delta P \equiv P(T,\mu) - P(T, \mu=0)$).  
Here $\mu$ is the quark chemical potential which is one 
third of the baryon chemical potential. The present LGT results are 
with $\mu_B$ coming from $up$ quarks only and hence we need to consider 
only $up$ quark density.     
Using the expression for the number density, Eq. (\ref{eq:n}), 
which on continum limit and after some algebra, reduces to
\begin{equation}
 \frac{n_q}{T^3} = \frac{12}{\pi ^2} \sum_{l=1}^\infty
(-1)^{l-1} \frac{1}{l^3} \left[ (\frac{m_q\,l}{T})^2 K_2 (\frac{m_q\,l}{T})
\sinh(\frac{\mu \,l}{T} \right]
\,\,. \label{eq:nq} \end{equation}
Now we modify earlier $m_{th}^2 (T)$ to $m_{th}^2 (T,\mu)$ as
\begin{equation}
m_{th}^2 (T,\mu) = \frac{g^2 T^2}{18} n_f ( 1 + \frac{\mu}{\pi^2\, T^2})
\,\,, \end{equation}
inspired by QCD perturbative calculations \cite{pe.1}.
In our case $n_f = 2$, mainly due to $up$ quarks, and $g^2$
is related to two-loop order running coupling constant, discussed earlier,
but need to be modified to take account of finite $\mu$. Following the work of
Schneider \cite{sch.1} and Letessier, Rafelski \cite{ra.1},
now we change $T/\Lambda_T$ in Eq. (\ref{eq:ls}) as
\begin{equation}
\frac{T}{\Lambda_T} \,\sqrt{1 + a \,\frac{\mu^2}{T^2} } \,\,,
\label{eq:lsa} \end{equation}
where $a$ is a parameter which is equal to $(1.91/2.91)^2$ in the
calculation of Schneider 
and $1/\pi^2$ in a phenomological model of Letessier and Rafelski. 
In our model Schneider's $\alpha_s (T,\mu)$ works well. 
                                                                            
From $n_q$, we may obtain other thermodynamic quantities like,
\begin{equation}
\Delta P \equiv P(T,\mu) - P(T,0) = \int_0^\mu n_q d\mu \,\,, 
\end{equation}
and so on. 
 
\section{Results :} 

In Fig. 1, we plotted $P/T^4$ Vs $T/T_c$ for (2+1)-flavor QGP with 
$\mu = 0$ and compared with LGT data. We took effective number of flavors 
as $n_u^{eff} = 2$, $n_s^{eff} = .5$ and $n_f^{eff} = 2.5$ because of finite 
quark masses. Quark rest masses are $m_{u0} = 65\,MeV$ and $m_{s0} = 165\,MeV$, values used in LGT simulations.  
Only the parameter $t_0 \equiv \Lambda_T /T_c$ is 
adjusted to get the best fit to LGT data on pressure and is equal 
to $0.4$. In Fig. 2, the baryon density ($n_B /T^3$) is plotted as a function 
of $T/T_c$ for different values of baryon chemical potentials and compared with
LGT results. Surprisingly good fit is obtained without any new parameters. 
In Fig. 3, $\Delta P / T^4$ is plotted and again very good fit to LGT. 

\section{Conclusions :} 

Using our new formulation of qQGP phenomenological model, we were able 
to explain LGT results on (2+1)-flavor QGP with just a single parameter 
which may be related to QCD scale parameter. Our formalism is 
thermodynamically consistent and no need of reformulation of SM with 
extra temperature dependent terms in pressure and energy density so on. 
Earlier, using this model, we explained successfully the LGT results of 
Bielefeld group on QGP with massless quarks \cite{ba.2}. 
Hence a simple model of qQGP with a single parameter, related to QCD 
scale parameter, explains all existing LGT results and may explain 
future results in LGT and relativistic heavy ion collisions. 

Of course, there are many other models which all claims to fit the LGT 
results, but all of them involve more than one parameter. Models 
with minimum number of parameters, which fits LGT results well, are 
SCQGP \cite{ba.3} and liquid model \cite{ra.2,ra.1}, both with two parameters. 
All other models involve more than two parameters. Models based on 
QCD perturbative and non-perturbative calculations \cite{bi.1,d.1} 
fails to fit the LGT results. Hence, it seems, phenomenological models 
based on the properties of plasma with QCD inputs explains well the 
LGT results.

\newpage
\begin{figure}
\caption { Plots of $P/ T^4 $ as a function of $T/T_c$ from 
our model and lattice results (symbols) \cite{fk.1} for (2+1)-flavor QGP. } 
\label{fig 1}
\vspace{.75cm}

\caption { Plots of $n_B/T^3$ as a function of $T/T_c$ from 
our model of (2+1)-flavor QGP and also 
with lattice data (symbols) \cite{fk.1} for $\mu_b =$ $100$, $210$, $330$, 
$410$ and $530$ $MeV$ from bottom to top. } 
\label{fig 2}
\vspace{.75cm}

\caption { Plots of $\Delta P/T^4$ as a function of $T/T_c$ from 
our model of (2+1)-flavor QGP and also 
with lattice data (symbols) \cite{fk.1} for $\mu_b =$ $100$, $210$, $330$, 
$410$ and $530$ $MeV$ from bottom to top. } 
\label{fig 3}
\vspace{.75cm}

\end{figure}
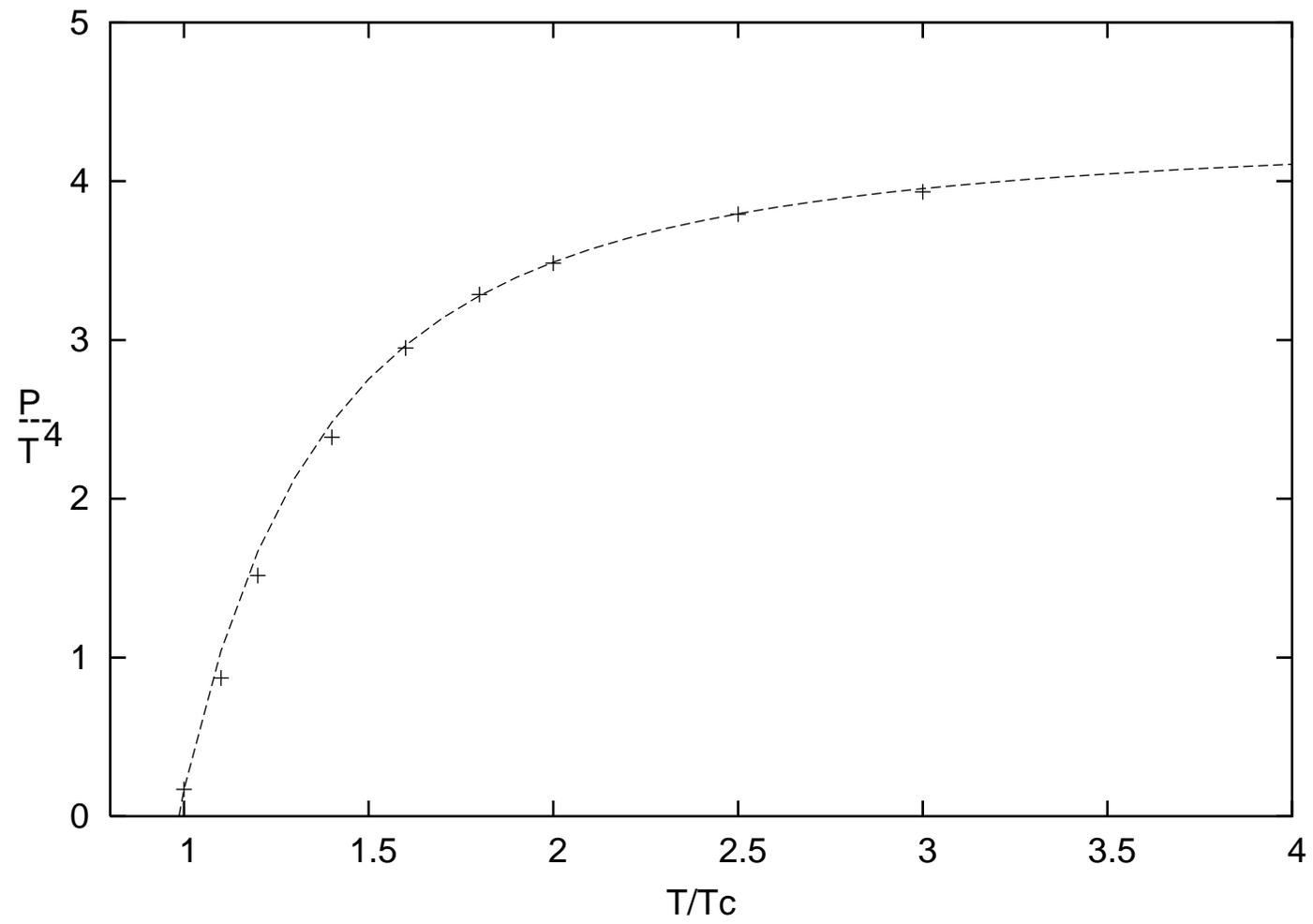
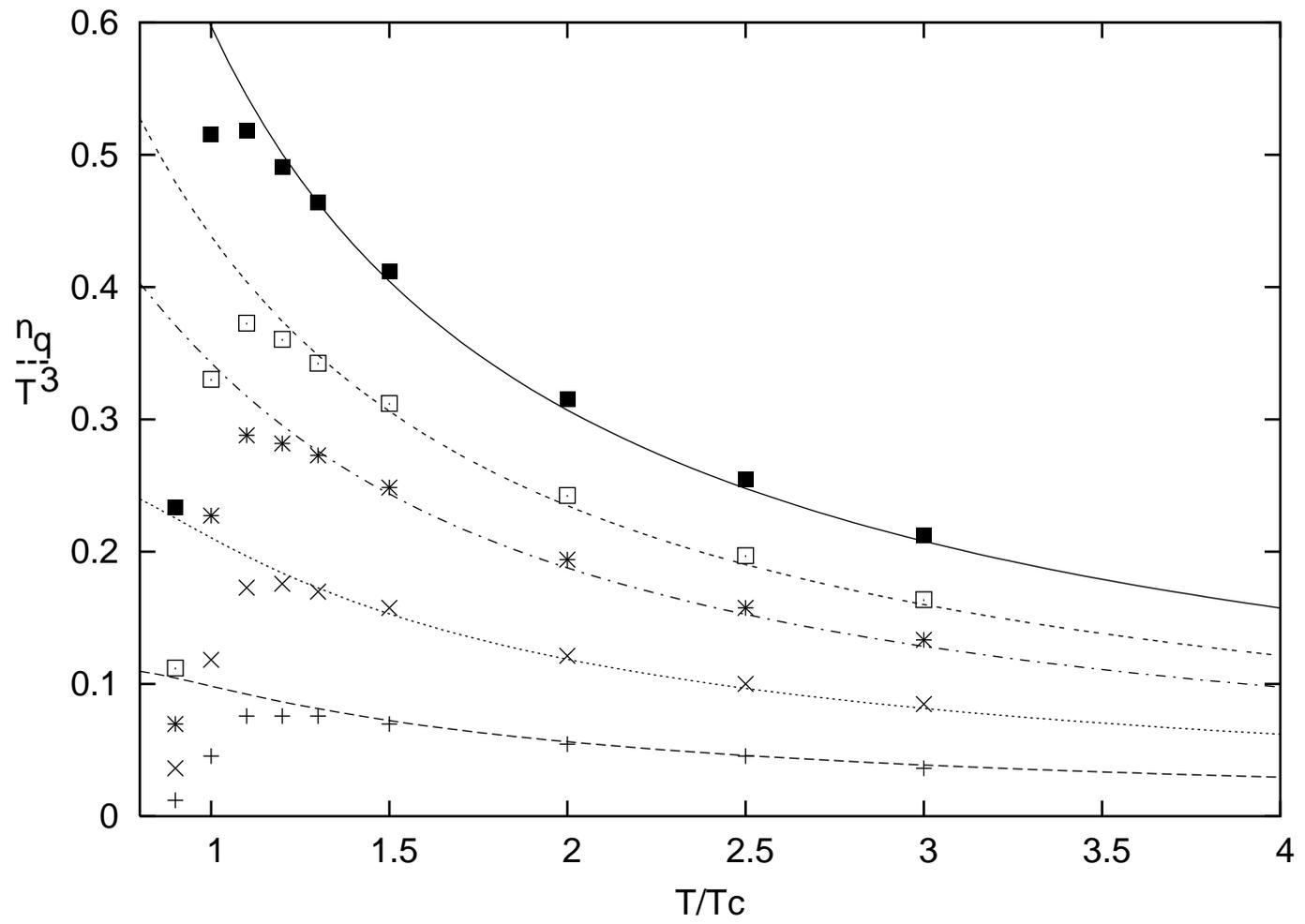
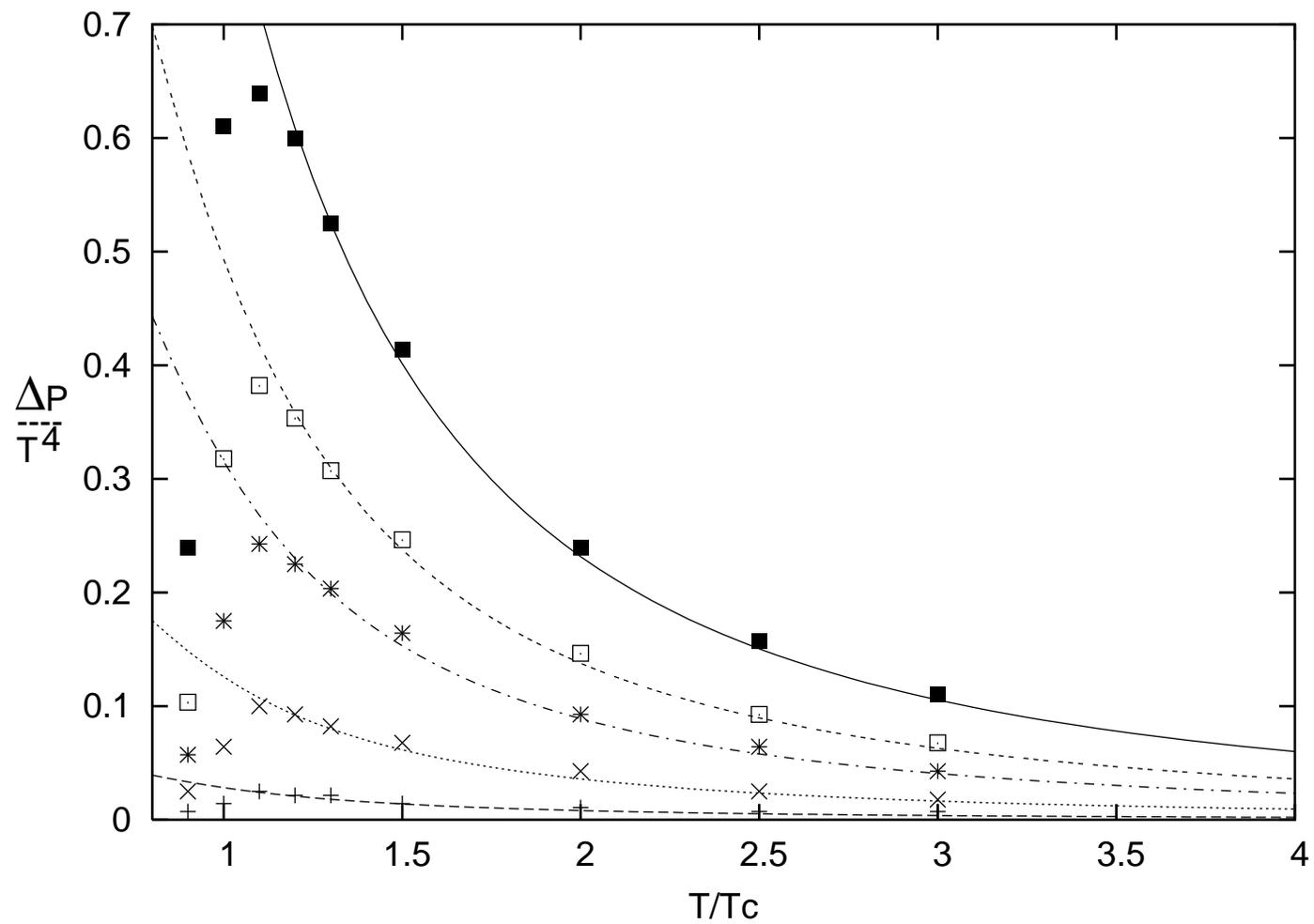
\end{document}